\documentclass[showpacs,preprintnumbers,amsmath,amssymb,twocolumn]{revtex4}
\usepackage{graphicx}
\usepackage{dcolumn}
\usepackage{bm}
\begin{document}
\newcommand{\vs}{}
\title{Novel Dynamics and Thermodynamics in systems with long range interactions}
 \author{Boris Atenas and Sergio Curilef}
\affiliation{%
Departamento de F\'\i sica, Universidad Cat\'olica del Norte,
Av. Angamos 0610, Antofagasta, Chile.}
\date{\today}
\begin{abstract}
 Systems with long range interactions display some anomalies when its dynamics and thermodynamics are studied below certain conditions.
 Among these anomalies are the quasi-stationary states, which are exacerbated because of special initial conditions that are used here. We present in this letter a new Hamiltonian whose potential is inspired in the two-dipole interaction. An analytical solution is obtained for the equilibrium in the canonical ensemble that is coincident with the one obtained from computational simulations. However, results from this model presents a kind of nonequivalence of ensembles in long-living states before arriving to equilibrium. Thus, a complete characterization is made for the nonequilibrium through molecular dynamics. In which, novel quasi-stationary states are observed due to the long range interactions.
\end{abstract}

\maketitle 

From years ago, the Ising model has been considered the most relevant tool                                                                            to study magnetic properties and the statistical behavior of many-body systems. Extra efforts have been made to propose several variations of the Ising model, specially for theoretical and numerical modeling of systems with long range interactions.

Typically, it has been accepted that at low energy in this kind of systems arises a phase identified by the presence of a single cluster of particles floating in a diluted homogeneous background. At high energy a homogeneous phase is recovered; the cluster disappears and the particles move almost freely. In the transition region, the system is characterized by the microcanonical ensemble with a negative specific heat: the resulting instability is extremely relevant[] because of its strong implications on experimental and theoretical features. This corresponds to an apparent thermodynamical inconsistency, which has been solved by Hertel and Thirring\cite{HertelAOP1970}. They proposed that the canonical and microcanonical ensembles are not equivalent close to the transition region. Until now, this proposal has been successfully confirmed by numerical simulations of systems with long range interactions\cite{pre70,mplb17,pra45,prl76,prl86}.

Additionally, specialists have been developing various methods \cite{Ribeiro,Buyl,Ruth,Yoshida} to manipulate in computational simulations and theoretical descriptions adequately the interactions in order to obtain a proper characterization of the behavior of these systems because of the wide range of applications that we find in nature that going from microscopic to macroscopic scale. For instance, the possibility of controlling the matter at the molecular level to develop nano-machinery is a dream from decades ago. In the other limiting case, the possible understanding of the behavior in the astronomical scales is other motivation to catch deeply this kind of systems. Certainly the systems in nature are idealized without interactions, but a deeper understanding needs to include interactions, which can be extracted from this class of pivotal descriptions.

Several mean field models have been recently introduced to study these anomalies related to long-range interactions.
In part, the behavior of the kinetic energy and other thermodynamic observables are used to characterize the stationary states, where long-living states before arriving to equilibrium are observed\cite{Lyndenbell,{Pluchino},Campa}. One of these intriguing models is the Hamiltonian mean field model, whose properties become paradigmatic and pertinent to characterize general systems with long range interactions. However, this behavior seems to be persistent and it has been observed in several self-gravitating models and other systems with long range interactions.

The main goal in this work is to study anomalous behavior of systems with long range interactions from a new Hamiltonian mean field model inspired in the two-dipole interactions. We introduce this model increasing the stock of mean field models. We expect to bring other nonstandard properties and to correctly characterize them. The dependence on the orientations of the dipoles is characterized by the zeroth-order approximation of the dipolar interaction. In this work, we use the canonical ensemble to calculate the free energy, magnetization and internal energy for the equilibrium. Through numerical simulation we characterize the mean kinetic energy, distributions, caloric curve, mean square of the displacement to obtain the diffusion law. We observe that due to out of equilibrium states, we can define two different quasi-stationary states (QSS). This anomaly is a novel observation in comparison with other models known in literature.

Now, for theoretical and numerical modeling of systems with long range interactions, we take a system of N identical coupled particles, with  mass equal to 1,  whose dynamics evolves in a periodic cell described by a 1D Hamiltonian given by
\begin{equation}\label{powerlaw}
 H \!= \!\sum_{i=1}^{N}\!\frac{p_i^2}{2}\!+\! \frac{\epsilon}{2N} \!\sum_{i\neq j}^N [\cos(\theta_i\!-\!\theta_j)\!-\!3\cos\theta_i \cos\theta_j\!-\!\Delta_{i,j}]
\end{equation}
where $p_i$ and $\theta_i$ represent the momentum and the angle of orientation of the particle $i$, with $i=1,\cdots N$, being $N$ the size of the system. The parameters $\epsilon$ and  $\Delta_{i,j}$ stand for the coupling and initial conditions. Meanwhile, the parameter $\Delta_{i,j}$ is defined to consider properly the zero of the energy as follows:
\begin{equation}
\Delta_{i,j}=\cos(\theta_{0i}\!-\!\theta_{0j})\!-\!3\cos\theta_{0i} \cos\theta_{0j},
\end{equation}
thus, the set of angles $\{\theta_{0k}\}$ stands for the initial orientations of the dipoles.
The interaction coupling is rescaled by the number of particles to make the potential thermodynamically stable[]. If $\epsilon$ is positive, the system is ferromagnetic, but if $\epsilon$ is negative the system is antiferromagnetic.
The equilibrium state can be exactly derived, however the QSS are not standard and cannot be exactly derived. Complementary to this, the implementation of numerical methods comes being an acceptable tool at the moment of studying and characterizing these anomalies.

The spin vector related to each particle is given by
\begin{equation}\label{spinvector}
\overrightarrow{m}_i=(\cos\theta_i,\sin\theta_i)
\end{equation}
Therefore, we can introduce the total spin vector
\begin{equation}\label{totalspin}
\overrightarrow{M}=\frac{1}{N}\sum_{i=1}^N\overrightarrow{m}_i=(M_x,M_y) = M\exp(i\phi)
\end{equation}
where $(M_x,M_y)$ and $M$ are the components and the modulus of the vector $\overrightarrow{M}$, respectively, $\phi$ stands for the phase of the order parameter. The equation of motion is
\begin{equation}\label{eqmot}
\dot{p}_i=-\frac{\epsilon}{2N}\left(2M_x\sin\theta_i +M_y\cos\theta_i\right)
\end{equation}
and the potential can be written as follows
\begin{equation}
V =  \frac{\epsilon}{2N} (2M_x^2-M_y^2-\Delta),
\end{equation}
where $\Delta =\sum_{i,j}\Delta_{i,j}$. In the canonical ensemble, the partition function
\begin{equation}
Z(\beta,N)\!=\!\!\int \texttt{d}^N\!\!p_i\, \texttt{d}^N\!\!\theta_i e^{-\beta N}=Z_K(\beta,N)Z_V(\beta,N),
\end{equation}
where $Z_K(\beta,N)$ is the kinetic part of the integral and the $Z_V(\beta,N)$ the interacting part. Therefore,

\begin{equation}
Z_K(\beta,N)=\int \texttt{d}^N\!\!p_i  \exp\left(-\frac{\beta}{2} \sum_i p_i^2\right)=\left(\frac{2\pi}{\beta}\right)^{N/2}
\end{equation}
In addition, considering the Eqs.(\ref{spinvector}) and (\ref{totalspin}) we can write
\begin{equation}
Z_V(\beta,N)=\int \texttt{d}^N\!\!\theta_i \exp\left(-\frac{\beta\epsilon}{2N} (2M_x^2-M_y^2-2)\right)
\end{equation}
 if $\{\theta_{0,i}=0 \}$ for all $i$, $\Delta =-2$. Now, taking into account, both the real and complex Hubbard-Stratonovich transformations\cite{Stratonovich1,Stratonovich2}, we obtain
\begin{equation}
Z_V(\beta,N)=e^{-\beta \epsilon N}\int_{-\infty}^\infty \texttt{d}x\, e^{-\beta \epsilon N x^2} \textrm{I}_0(2\beta \epsilon x)^N
\end{equation}
where $\textrm{I}_k(y)$ is the modified Bessel function of k\emph{th}-order. This last integral can be evaluated through the saddle point method in the thermodynamic limit, $N\rightarrow \infty$. The free energy per particle $\varphi$ is given by
\begin{equation}
\varphi(\beta,N)\!=\!\frac{1}{2}\ln\frac{\beta}{2\pi} \!+\!\epsilon\beta \!+\!\inf_{x\geq 0}\![-\!\beta\epsilon x^2\!+\!\ln \textrm{I}_0(2\beta\epsilon x)]
\end{equation}
The solution of the extremal is obtained by the transcendental equation,
\begin{equation}
x =\frac{\textrm{I}_1(2\beta \epsilon x)}{\texttt{I}_0(2\beta \epsilon x)}.
\end{equation}
The procedure is particular to this problem, but in part analogous to the Hamiltonian mean field model. The critical temperature is $T_c=1$, that is the double compare with the $T_c$ obtained for the Hamiltonian mean field model. The clustered phase is found for $T<T_c$ and the homogeneous phase occurs for $T>T_c$.
\begin{figure}[t!]
\centering
\includegraphics[width=1\linewidth]{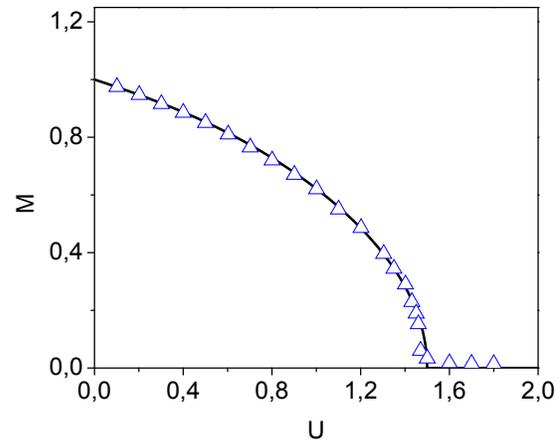}
\caption{Equilibrium magnetization as a function of internal energy $U$. Triangles are data obtained from the simulation. The solid line is the analytical solution from canonical ensemble: $U=\frac{1}{2\beta}+1-M^2$.}\label{Mag1}
\end{figure}
If $\epsilon <0$, the equation has a trivial solution, $x=0$. By contrary, if $\epsilon >0$, the equation has a set of values for $x$ and $\beta$, which defines the solution of the problem. Finally, it is obtained the internal energy per particle, as a function of the temperature and magnetization, as
\begin{equation}\label{ASolution}
U=\frac{\partial \varphi(\beta,N)}{\partial \beta} = \frac{1}{2\beta}+1-M^2
\end{equation}
where M, is the solution of the extremal problem.

Numerical simulations are carried out, by microcanonical molecular dynamics, to check the validity of the analytical results, by the canonical ensemble. Solving numerically the equations of motions considering the special initial condition called \emph{water bag initial conditions} (WBIC), we show that the equilibrium is well described by the canonical ensemble.  The magnetization $M$ is the clustering degree of the particles. It is possible to define the critical temperature $T_c$ where the continuous trend of $M$ vanishes.

The magnetization $M$ is obtained from the Eq.(\ref{ASolution}) as a function of the internal energy $U$, which corresponds to the solution we derive from the canonical ensemble. Correspondingly, we obtain solution by simulations in the microcanonical ensemble for several energies.
In Fig.\ref{Mag1} we depict the equilibrium magnetization $M$ as a function of the internal energy $U$. Numerical data are represented by triangles. The analytical solution is depicted by solid line from canonical ensemble given by Eq.(\ref{ASolution}). The critical point is located at $U_c=3/2$ just twice the value obtained for the Hamiltonian mean field model.

\begin{figure}[t!]
\centering
\includegraphics[width=1\linewidth]{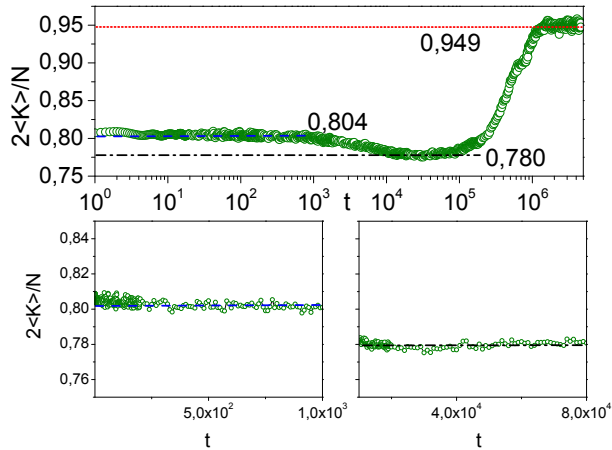}
\caption{In the figure it is depicted the evolution of the kinetic energy, it is shown two QSS previous to the equilibrium. The first appears for \:\:\, $10<t<10^3$ mean kinetic energy $2<K>/N_{QSS1}=0.775$ (dashed line). The second occurs for $10^4<t<4\cdot10^5$ with $2<K>/N_{QSS2}=0.761$ (dashed-dotted line) and finally $T_{eq}=0.948$ (dotted line). The data are the mean value of 100 samples with $N=8192$ and $U=1.38$.}\label{qss}
\end{figure}
Other challenge to characterize the current model is to evaluate the kinetic energy. The behavior of this thermodynamic quantity is relevant to define properties of the system. In the dynamics of the model is it is important to observe where the kinetic energy is constant. As far as we know, this is relevant to search the behavior of other thermodynamic observables.
In Fig.\ref{qss} we represent the dynamics of the system by the mean value of the kinetic energy in $100$ samples with the size $N=8.192$ particles. As said before, regions where the kinetic energy is stable are interesting for the present study. First ($10 < t < 10^3$) and second ($10^4 < t < 4\times \,\, 10^5$) regions define the called QSS. In this model we clearly identify two time intervals where the kinetic energy is constant, previous the last one where the equilibrium is reached. The value of first stable kinetic energy is upper than the second value.
We use $U=1.38$ because the system has the maximum anomaly when the stability of the kinetic energy is observed. For this value of energy $U$, the temperature  that is related with the kinetic energy in the equilibrium, it is obtained analytically $T_{eq}=0.948$ and coincides with the numerical data with 0.3$\%$ of exactness.

\begin{figure}[t!]
\centering
\includegraphics[width=1\linewidth]{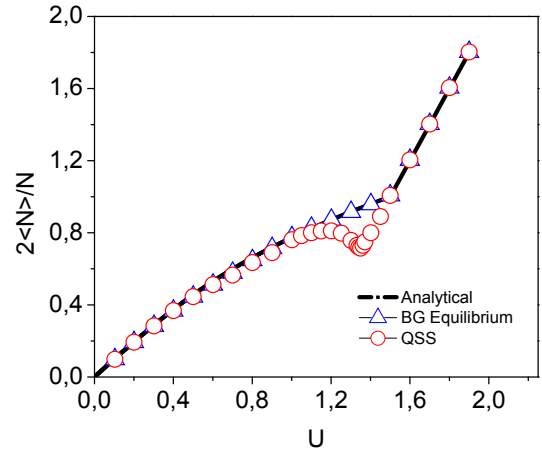}
\caption{It is depicted the $2<K>/N_{QSS2}$ for QSS$_2$ (circles) and its correspondingly equilibrium values (triangles) as a function of the internal energy $U$. In consequence, equilibrium values match the $T_{eq}$ (solid line), which coincide with the analytical solution in the canonical ensemble. Simulation takes $N=4096$. }\label{TvsU}
\end{figure}

After observing the difference among the value of the kinetic energy in equilibrium with the corresponding values out of equilibrium, these are the two QSS, we choose the deepest numerical result to compare with the equilibrium that we obtain from two ways, the theoretical canonical ensemble and the numerical microcanonical simulation. Therefore, in the Fig.\ref{TvsU} we superpose three curves that represent twice kinetic energy per particle $2<K>/N_{QSS2}$ as a function of internal energy $U$. Triangles stand for equilibrium data (from the numerical simulations) that coincide exactly with the analytical solution. But, circles correspond to the QSS in the second region. Here, we notice that there is a region where one set of values is not coincident with the other. In particular, there is a point, specifically $U=1.38$ where the difference is the greatest. This point has been used to characterize the QSS. In the interval from 1.1 to 1.38 we observe a disagreement between molecular dynamics simulation and canonical ensemble treatment in the limit o the mean field approximation.

\begin{figure}[t!]
\centering
\includegraphics[width=0.5\textwidth]{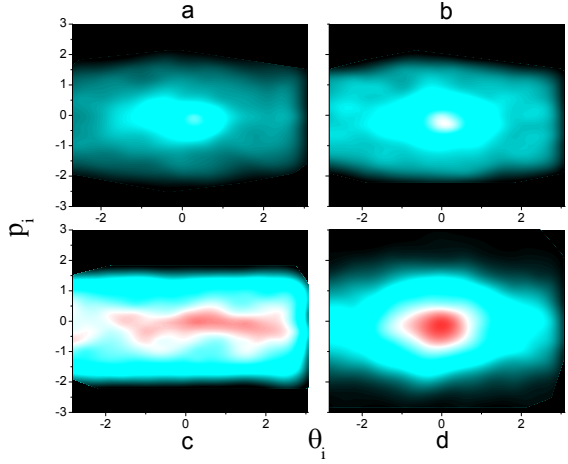}
\caption{Snapshots of the distribution function for the dipolar Hamiltonian mean field model into phase space. (\emph{a}) For $t=20$ weakly deviated from WBIC. (\textbf{b}) For $t=200$, the system is in the first region of QSS. (\emph{c}) For $t=600.000$, the system is in the second region of QSS. Finaly (\emph{d}) For $t=5.000.000$, the system reaches the equilibrium where ellipses characterized trajectories into phase space.}\label{F4}
\end{figure}

In order to fully characterize the dynamics of the system, we are interested in observing the behavior in phase space. Thus, for instance, we can take some pictures at several states of the evolution. Thus, in Fig.\ref{F4} it is shown some snapshots at different times to characterize the distribution in phase space; namely, $t=20$ close to WBIC, at $t=200$ the first region of QSS, at $t=600.000$ the second region of QSS and at $t= 5.000.000$ the equilibrium. In Fig.\ref{F4} (\emph{a}), we see the distribution slightly deformed from the WBIC with defined regions occupying the phase space. In Fig.\ref{F4}(\emph{b}), the distribution is extended by all phase space in a irregular form. In Fig.\ref{F4}(\emph{c}), the distribution takes paths enough defined showing certain regularities that we see in the representation into phase space. In addition, we see in Fig.\ref{F4}(\emph{d}) trajectories that are defined by elliptical shapes represented by several colors.

\begin{figure}[h!]
\centering
\includegraphics[width=0.5\textwidth]{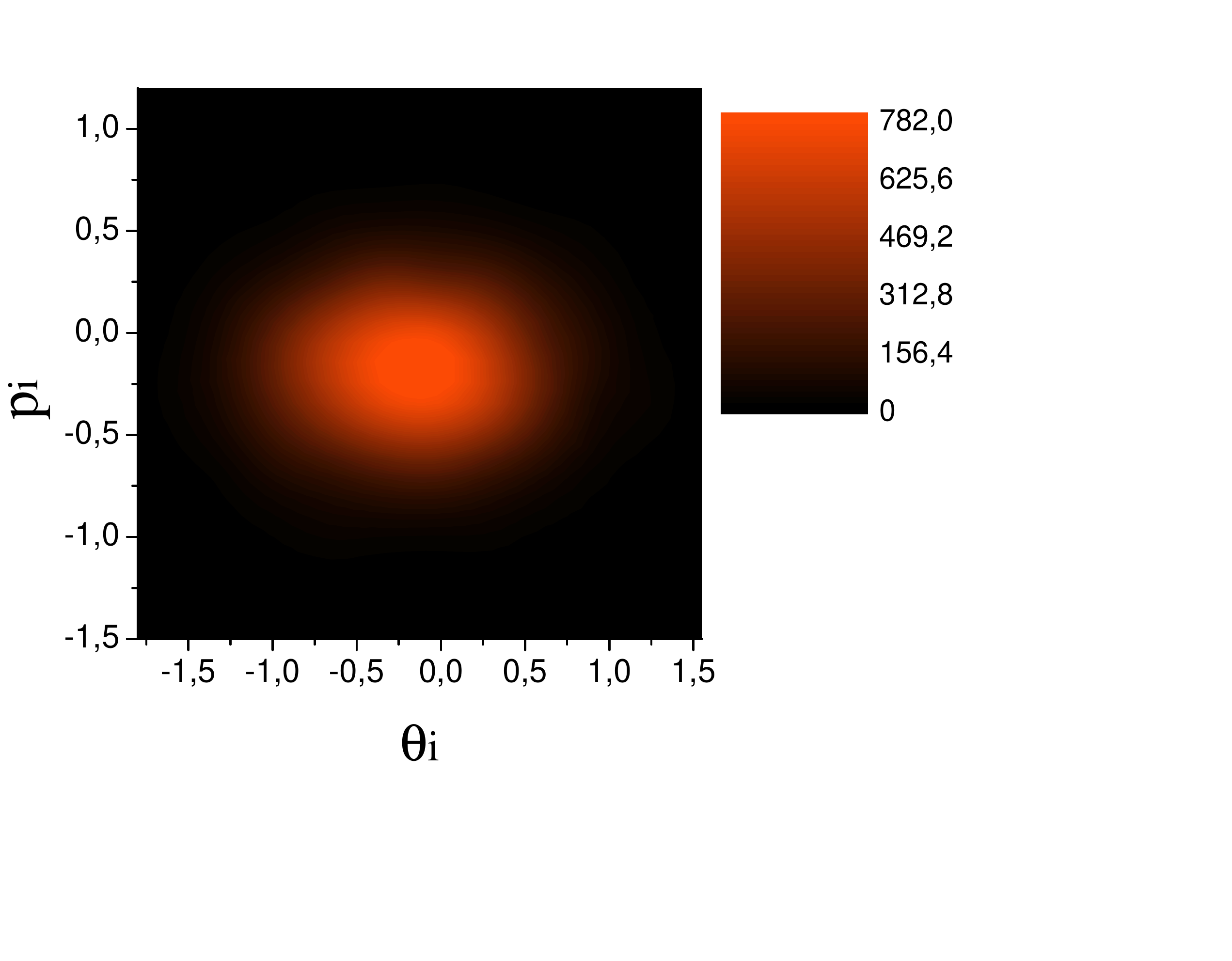}
\caption{Equilibrium distribution is depicted in phase space.}\label{F5}
\end{figure}

In addition, in Fig.\ref{F5} it is represented a typical equilibrium distribution in phase space, obtained from the average of among various snapshots in similar conditions that we show in Fig.\ref{F4}(d).

\begin{figure}[t!]
\centering
\includegraphics[width=0.5\textwidth]{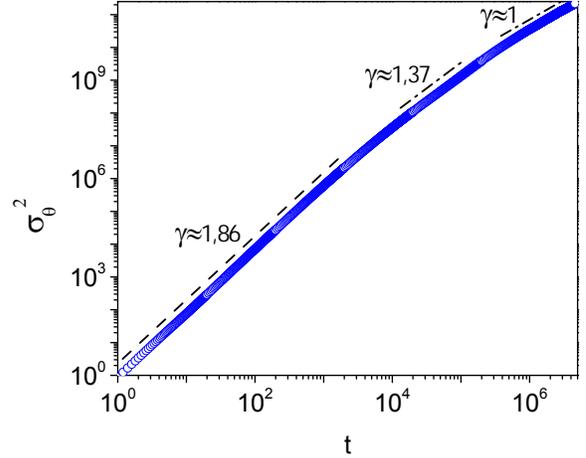}
\caption{The dynamics is illustrated by the evolution of the variance. Anomalous   diffusion is obtained in two regions that coincide with the two QSS previously discussed. The equilibrium is obtained when $\gamma=1$.}\label{F6}
\end{figure}

Finally, we represent the dynamics of particles of the system, by the variance of the angular displacement defined as $\sigma^{2}_\theta (t) =\sum_i(\theta_i(t)-\theta_i(0))^2/N$. Equations of motion are obtained from the Eq.(\ref{eqmot}) that are numerically integrated. We choose a combination of parameters where it is possible to observe dynamics at initial conditions out of equilibrium and that progressively acquires states with anomalous diffusion until the equilibrium is reached. Specifically, two superdiffusive states are observe in Fig.\ref{F6}, where $\sigma_\theta^2 \propto t^\gamma$  and   $\gamma > 1$.

We would like to acknowledge partial financial support by CONICYT-UCN PS-065. One of us (B.A.) for the Beca de Magister Concurso Nacional 2014 del Conicyt (Graduate fellowship by Conicyt 2014). We appreciate the computational assistance of A. Pluchino.


\newpage
\begin{thebibliography}{88}
\bibitem{HertelAOP1970} P. Hertel and W. Thirring, Ann. Phys. \textbf{63}, 520 (1970)
\bibitem{pre70} H. Chamati, D. Dantchev, Phys. Rev. {\bf E 70}, 066106 (2004).
\bibitem{mplb17} H. Chamati, N. Stonchev, Mod. Phys. Lett. {\bf 17}, 1187 (2003); J. Phys. A: Math. Gen 33, L187 (2000).
\bibitem{pra45} R. Minieri, Phys. Rev. {\bf A 45} 3580 (1992).
\bibitem{prl76} E. Luijten and H. W. J. Bl\"ote, Phys. Rev. Lett. {\bf 76}, 1557 (1996).
\bibitem{prl86} E. Luijten and H. Me$\beta$ingfeld, Phys. Rev. Lett. {\bf 86}, 5305 (1996).
\bibitem{AntoniPRE57} M. Antoni and A. Torcini, Phys. Rev. \textbf{E 57}, 6233 (1998)
\bibitem{Curilef} S. Curilef, L. A. del Pino and P. Orellana, Phys. Rev. {\bf B 72}, 224410 (2005)
\bibitem{kac} M. Kac, G. Uhlenbeck, P. C. Hemmer, J. Math. Phys. \textbf{4}, 216 (1963)
\bibitem{Lyndenbell}  D. Lyndenbell, 
Monthly Notices of the Royal Astronomical Society \textbf{136}, 101 (1967)
\bibitem{Pluchino} A. Pluchino, V. Latora and A. Rapisarda, 
Springer-Verlag, Volume 16, Issue 3, pp 245-255 (2004).
\bibitem{Campa} A. Campa, T. Dauxois and S. Ruffo, 
Physics Reports 480, 57-159 (2009).
\bibitem{PhysA340} S. Curilef, Physica \textbf{A 344}, 456 (2004)
\bibitem{delPinoPRB2007} L. A. del Pino, P. Troncoso and S. Curilef, Phys. Rev. \textbf{B 76}, 172402 (2007)
\bibitem{AtenasAOP2014} B. Atenas, L. A. del Pino and S. Curilef, Annals of Physics \textbf{350}, 605-614 (2014)
\bibitem{Ribeiro} A. C. Ribeiro-Teixeira, F. P. C. Benetti, R. Pakter, and Y. Levin, Pysical Review E 89, 022130 (2014)
\bibitem{Buyl} Pierre de Buyl, Computer Physics Communications 185 (2014) 1822"1¤71827.
\bibitem{Ruth} R. Ruth and E. Forest, 
Physica \textbf{D 43}, 105-117 (1990).
\bibitem{Yoshida} H. Yoshida, 
Physics Letters \textbf{A 150}, 262 (1990)
\bibitem{Stratonovich1} R. L. Stratonovich, Soviet Physics Doklady \textbf{2}, 461 (1958)
\bibitem{Stratonovich2} J. Hubbard, Phys. Rev. Lett. \textbf{3}, 77"1¤778 (1959)
\end{thebibliography}
\end{document}